\documentclass[prd,amsmath,amssymb,longbibliography,superscriptaddress,twocolumn,nofootinbib,10pt]{revtex4-1}

\pdfoutput=1

\usepackage{graphicx}
\usepackage{dcolumn}
\usepackage{bm}
\usepackage{amssymb}
\usepackage{latexsym}
\usepackage{booktabs}
\usepackage{amsmath}
\usepackage{multirow}
\usepackage[colorlinks=true, linkcolor=blue, citecolor=blue]{hyperref}

\begin{document}

\title{How Dark Sector Equations of State Govern Interaction Signatures}

\author{Peng-Ju Wu}\thanks{Corresponding author}\email{wupengju@nxu.edu.cn}
\affiliation{School of Physics, Ningxia University, Yinchuan 750021, China}

\author{Ming Zhang}
\affiliation{School of Mathematics and Physics, Qingdao University of Science and Technology, Qingdao 266061, China}

\author{Shang-Jie Jin}
\affiliation{Liaoning Key Laboratory of Cosmology and Astrophysics, College of Sciences, Northeastern University, Shenyang 110819, China}
\affiliation{Department of Physics, University of Western Australia, Perth WA 6009, Australia}
\affiliation{Research Center for the Early Universe, Graduate School of Science, The University of Tokyo, Tokyo 113-0033, Japan}

\begin{abstract}
Using late-Universe observations, we demonstrate that freeing dark energy and dark matter equations of state (EoS) dramatically alters the inferred strength and direction of their interactions. When dark sector EoS are fixed to $w_{\mathrm{de}}=-1$ and $w_{\mathrm{dm}}=0$, the data consistently favor an energy transfer from dark energy to dark matter across various interaction forms. This apparent evidence, however, proves highly sensitive to the EoS assumptions: treating $w_{\mathrm{de}}$ as a free parameter substantially weakens the evidence for interaction, with its value converging to the quintessence regime ($w_{\mathrm{de}}>-1$). In contrast, freeing $w_{\mathrm{dm}}$ maintains a preference for interaction, revealing a correlation where positive $w_{\mathrm{dm}}$ is associated with energy transfer from dark energy to dark matter, and negative $w_{\mathrm{dm}}$ with energy transfer from dark matter to dark energy. These findings caution against the simplistic assumption of $\Lambda$CDM EoS values when attempting to detect a possible interaction. Despite these fundamental degeneracies, model comparison indicates that interacting dark energy scenarios are positively to strongly supported by AIC and DIC, but only inconclusively to weakly supported by Bayesian evidence against the $\Lambda$CDM model.
\end{abstract}

\maketitle
\section{Introduction}\label{sec1}
The $\Lambda$CDM paradigm, which synthesizes a cosmological constant $\Lambda$ with cold dark matter and standard baryonic physics, provides an economical fit to a wide range of astronomical observations, from the cosmic microwave background (CMB) anisotropies to the large-scale clustering of galaxies \cite{Planck:2018vyg,eBOSS:2020yzd}. Yet the origin of either dark sector remains unknown. In $\Lambda$CDM, dark energy is treated as a perfect fluid with equation of state (EoS) parameter $w_{\rm de}=-1$ at all epochs, while dark matter is assumed to be collisionless and non-relativistic with $w_{\rm dm}=0$. These two components are postulated to evolve independently. However, given their unknown nature, the possibility of a direct interaction between them remains a theoretically compelling extension, prompting investigations into models beyond the minimal standard paradigm \cite{Amendola:1999er,Farrar:2003uw,Valiviita:2008iv,Li:2014eha,Li:2013bya,Wang:2016lxa,Wang:2024vmw,vanderWesthuizen:2025vcb,vanderWesthuizen:2025mnw,vanderWesthuizen:2025rip}.

The phenomenological foundation of interacting dark energy (IDE) models is characterized by a set of coupled conservation equations. In the background, this interaction is typically encoded through a source term $Q$ in the continuity equations:
\begin{align}\label{IDEQ}
\dot{\rho}_{\rm de}+3H(1+w_{\rm de}){\rho}_{\rm de}=+Q, \nonumber \\
\dot{\rho}_{\rm dm}+3H(1+w_{\rm dm}){\rho}_{\rm dm}=-Q,
\end{align}
where $\rho_{\rm de}$ and $\rho_{\rm dm}$ represent the energy densities of dark energy and dark matter, respectively, and the dot denotes the derivative with respect to cosmic time. The source term $Q$ is usually taken as $Q=\beta H \rho_{\rm de}$ \cite{Billyard:2000bh,Amendola:1999qq}, where $H$ is the Hubble parameter and $\beta$ is a dimensionless coupling parameter; $\beta > 0$ means that dark matter decays into dark energy, $\beta<0$ indicates that dark energy decays into dark matter, and $\beta=0$ denotes that there is no interaction between them. The canonical IDE paradigm remains the $\Lambda$CDM values for the dark sector EoS. Under this minimal extension, analyses of observational data have revealed scenarios where a non-zero interaction term ($\beta \neq 0$) is statistically favored, suggesting the possible presence of an interaction between dark energy and dark matter \cite{Bertolami:2007zm,Salvatelli:2014zta,Ferreira:2014jhn,Yang:2018euj, DiValentino:2019ffd, DiValentino:2019jae,Nunes:2022bhn,Li:2024qso,Li:2025owk,Li:2025muv,Pan:2025qwy,Li:2015vla,Wolf:2025jed,You:2025uon,Wolf:2024stt,vanderWesthuizen:2025iam,vanderWesthuizen:2025iam,Paliathanasis:2025xxm}.

Dark energy may not be a mere cosmological constant and dark matter may not be strictly cold. With recent DESI releases, compelling evidence has emerged suggesting a statistically significant deviation of the dark energy EoS from $-1$ \cite{DESI:2024mwx,DESI:2025zgx,Scherer:2025esj,Sabogal:2025jbo,Wu:2025wyk,Colgain:2025nzf,Cai:2025mas}, alongside tantalizing hints of a non-zero EoS for dark matter \cite{Braglia:2025gdo,Li:2025dwz,Kumar:2025etf,Yang:2025ume,Chen:2025wwn,Wang:2025zri,Araya:2025rqz,Braglia:2025gdo}. Generally, a more negative dark energy EoS and an energy influx from dark matter to dark energy both enhance the observed cosmic acceleration; conversely, a positive dark matter EoS and an energy outflow from dark energy to dark matter produce analogous imprints on the expansion history. This physical mimicry inherently induces degeneracies between the interaction coupling parameter and both dark sector EoS \cite{Wang:2016lxa,Clemson:2011an,Pan:2022qrr,Shah:2025ayl}. A critical question thus arises: how would relaxing the assumptions on dark energy and dark matter EoS influence the direction and strength of the energy transfer in interactions under current observations? Given its importance, this paper undertakes an investigation of this issue.

Motivated by the distinct preferences between different observational datasets---for instance, where CMB data alone favor a coupling parameter $\beta < 0$, while combined CMB+DESI analysis prefers $\beta > 0$ (see Ref.~\cite{Li:2024qso} for details)---this work utilizes late-universe observations exclusively. This choice is further justified by the documented discrepancies in the measurements of key cosmological parameters like $H_0$, $S_8$, and $\Omega_K$ \cite{Verde:2019ivm,Riess:2021jrx,DiValentino:2019qzk,Handley:2019tkm,Wu:2024faw}, which often arise when comparing early and late-universe probes (though we note that different voices exist in the literature). By relying solely on late-universe observations, our analysis is designed to provide a complementary and self-consistent perspective on dark sector interactions, which is less susceptible to early-universe measurements.

\section{Method and Data}\label{sec2}
Previous studies have demonstrated that the strength and direction of the dark sector interaction are closely tied to its specific coupling form. To explore this dependence and to ensure the generality of our investigation, we consider three representative forms of the interaction term $Q$, where the energy transfer is proportional to the dark energy density, the dark matter density, and the sum of both dark sector densities, respectively \cite{Billyard:2000bh,Amendola:1999qq}. Specifically, we adopt the following three forms: $Q = \beta H \rho_{\text{de}}$, $Q = \beta H \rho_{\text{dm}}$, and $Q = \beta H (\rho_{\text{dm}} + \rho_{\text{de}})$.

For the coupling form $Q = \beta H \rho_{\text{de}}$, solutions to continuity equations are:
\begin{align}
\rho_{\text{de}}(a) &= \rho_{\text{de},0} a^{-3(1+w_{\text{de}}) + \beta},  \nonumber \\
\rho_{\text{dm}}(a) &= \rho_{\text{dm},0} a^{-3(1+w_{\text{dm}})} \nonumber \\
&- \frac{\beta \rho_{\text{de},0}}{\Delta} \left[a^{-3(1+w_{\text{de}}) + \beta} - a^{-3(1+w_{\text{dm}})}\right],
\end{align}
where $\Delta = 3(w_{\text{dm}} - w_{\text{de}}) + \beta$. For  the case $\Delta = 0$,
\begin{equation}
\rho_{\text{dm}}(a) = \rho_{\text{dm},0} a^{-3(1+w_{\text{dm}})} - \beta \rho_{\text{de},0} a^{-3(1+w_{\text{dm}})} \ln a.
\end{equation}
For the coupling form $Q = \beta H \rho_{\text{dm}}$, solutions to continuity equations are:
\begin{align}
\rho_{\text{dm}}(a) &= \rho_{\text{dm},0}a^{-3(1+w_{\text{dm}}) - \beta}, \nonumber \\
\rho_{\text{de}}(a) &= \rho_{\text{de},0} a^{-3(1+w_{\text{de}})}  \nonumber \\
&+ \frac{\beta \rho_{\text{dm},0}}{\Delta} \left[a^{-3(1+w_{\text{dm}}) - \beta} - a^{-3(1+w_{\text{de}})}\right],
\end{align}
where $\Delta = 3(w_{\text{de}} - w_{\text{dm}}) - \beta$. For the case $\Delta = 0$,
\begin{equation}
\rho_{\text{de}}(a) = \rho_{\text{de},0} a^{-3(1+w_{\text{de}})} + \beta \rho_{\text{dm},0} a^{-3(1+w_{\text{de}})} \ln a.
\end{equation}
For $Q = \beta H (\rho_{\text{dm}} + \rho_{\text{de}})$, the coupled system of differential equations is solved via eigenvalue decomposition. The continuity equations can be expressed in matrix form as:
\begin{equation}
\frac{\rm d}{{\rm d}\ln a}
\begin{pmatrix}
\rho_{\text{dm}} \\
\rho_{\text{de}}
\end{pmatrix}
= -\mathbf{A}
\begin{pmatrix}
\rho_{\text{dm}} \\
\rho_{\text{de}}
\end{pmatrix}.
\end{equation}
where the interaction matrix $\mathbf{A}$ is defined as:
\begin{equation}
\mathbf{A} =
\begin{pmatrix}
3(1+w_{\text{dm}}) + \beta & \beta \\
-\beta & 3(1+w_{\text{de}}) - \beta
\end{pmatrix}.
\end{equation}
The general solution for the dark sector energy densities is obtained by diagonalizing the matrix $\mathbf{A}$:
\begin{equation}
\begin{pmatrix} \rho_{\text{dm}} \\ \rho_{\text{de}} \end{pmatrix} =
C_1 \mathbf{v}_1 a^{-\lambda_1} + C_2 \mathbf{v}_2 a^{-\lambda_2},
\end{equation}
where $\lambda_1$ and $\lambda_2$ are the eigenvalues of matrix $\mathbf{A}$, $\mathbf{v}_1$ and $\mathbf{v}_2$ are the corresponding eigenvectors, $C_1$ and $C_2$ are constants determined by the initial conditions $\rho_{\text{dm}}(1) = \rho_{\text{dm},0}$ and $\rho_{\text{de}}(1) = \rho_{\text{de},0}$.

We constrain these IDE models using the late-time cosmological observations. The dataset comprises baryon acoustic oscillations (BAO), Type Ia supernovae (SN), cosmic chronometers (CC), strong lensing time delays (TD) and gamma ray bursts (GRB):
\begin{itemize}
    \item \textbf{BAO:} the DESI measurements of transverse, radial, and volume-averaged distance scales at $0.295<z<2.33$ \cite{DESI:2025zgx}.

    \item \textbf{SN:} 1829 distance moduli from DESY5 at $0.025<z<1.3$ \cite{DES:2024jxu}.

    \item \textbf{CC:} 32 Hubble parameter measurements at $0.07<z<1.965$ \cite{Moresco:2022phi}.

    \item \textbf{TD:} 7 time-delay distance and 5 angular diameter distance data at $0.295<z<0.745$ \cite{Wu:2025wyk}.

    \item \textbf{GRB:} 193 distance moduli calibrated via the Amati relation, spanning $0.034<z<8.1$ \cite{Amati:2018tso}.
\end{itemize}
These data may provide robust constraints through their complementary redshift coverage and sensitivity to different cosmological parameters. To derive constraints that are fully independent of the early-universe observations, we treat the sound horizon at the drag epoch $r_{\rm d}$ as a free parameter in the BAO analysis, rather than anchoring it to a value inferred from the CMB data. Additionally, we marginalize over the absolute magnitude $M_{B}$ of the SN Ia to account for the degeneracy with $H_0$.

It should be pointed out that despite substantial improvements in the quality and quantity of late-time observations, their constraining power remains limited for extensively extended parameter spaces. For this reason, we adopt a targeted strategy to mitigate parameter degeneracies. Rather than simultaneously freeing the EoS for both dark sectors, we examine the following three distinct scenarios:
\begin{itemize}
    \item Fixing dark energy and  dark matter EoS to their $\Lambda$CDM values ($w_{\text{de}} = -1$ and $w_{\text{dm}} = 0$).
    \item Freeing dark energy EoS while maintaining the cold dark matter assumption ($w_{\text{dm}} = 0$).
    \item Freeing dark matter EoS while maintaining the cosmological constant assumption ($w_{\text{de}} = -1$).
\end{itemize}
This approach allows us to probe possible deviations from the standard cosmological model in a controlled manner, providing clearer insights into how interactions manifest alongside non-standard dark sector properties.

\section{Results and Discussions}\label{sec3}
\begin{table*}[!htb]
\caption{Cosmological parameter constraints for the $\Lambda$CDM model, its extensions with a free dark energy or dark matter EoS, and IDE models using late-time observational data. All models are constrained with a combination of BAO, SN, CC, TD and GRB. Here $H_0$ is in units of $\rm km/s/Mpc$.}
\label{tab:results1}
\setlength{\tabcolsep}{2mm}
\renewcommand{\arraystretch}{1.4}
\begin{center}
\begin{tabular}{lcccccc}
\toprule[1.5pt]
\textbf{Interaction Form} & \textbf{Setup} & \textbf{$H_0$} & \textbf{$\Omega_{\rm m}$} & \textbf{$\beta$} & \textbf{EoS Parameter} \\
\midrule[0.8pt]

$Q = 0$
& $w_{\text{dm}}=0$; $w_{\text{de}}=-1$
& $73.57\pm0.87$                  & $0.3047\pm0.0075$                         & $0$                    & -- \\

$Q = 0$
& $w_{\text{dm}}=0$; Free $w_{\text{de}}$
& $72.40\pm0.88$                      & $0.2961\pm0.0090$                         & $0$                    & $w_{\text{de}} = -0.883^{+0.041}_{-0.037}$ \\

$Q = 0$
& Free $w_{\text{dm}}$; $w_{\text{de}}=-1$
& $72.80\pm0.90$                          & $0.365^{+0.024}_{-0.027}$                                & $0$                          & $w_{\text{dm}} = -0.059^{+0.026}_{-0.022}$ \\

\midrule[0.8pt]

$Q = \beta H \rho_{\text{de}}$
& $w_{\text{dm}}=0$; $w_{\text{de}}=-1$
& $72.51\pm0.91$                  & $0.378\pm0.024$                         & $-0.35\pm0.11$                    & -- \\

$Q = \beta H \rho_{\text{de}}$
& $w_{\text{dm}}=0$; Free $w_{\text{de}}$
& $72.50\pm0.91$                      & $0.275^{+0.210}_{-0.092}$                         & $-0.02^{+0.75}_{-0.38}$                    & $w_{\text{de}} = -0.89^{+0.25}_{-0.13}$ \\

$Q = \beta H \rho_{\text{de}}$
& Free $w_{\text{dm}}$; $w_{\text{de}}=-1$
& $72.40\pm0.92$                          & $0.372^{+0.021}_{-0.024}$                                & $-0.53^{+0.16}_{-0.39}$                          & $w_{\text{dm}} = 0.071^{+0.088}_{-0.100}$ \\

\midrule[0.8pt]

$Q = \beta H \rho_{\text{dm}}$
& $w_{\text{dm}}=0$; $w_{\text{de}}=-1$
& $72.78\pm0.91$                       & $0.347\pm0.018$                                     & $-0.180^{+0.074}_{-0.067}$                             & -- \\

$Q = \beta H \rho_{\text{dm}}$
& $w_{\text{dm}}=0$; Free $w_{\text{de}}$
& $72.33\pm0.91$                       & $0.239^{+0.043}_{-0.066}$                                     & $0.31^{+0.27}_{-0.35}$                                    & $w_{\text{de}} = -0.792^{+0.110}_{-0.063}$ \\

$Q = \beta H \rho_{\text{dm}}$
& Free $w_{\text{dm}}$; $w_{\text{de}}=-1$
& $72.81\pm0.89$                          & $0.51^{+0.11}_{-0.27}$                                     & $0.42^{+1.50}_{-0.54}$                                    & $w_{\text{dm}} = -0.20^{+0.20}_{-0.49}$ \\

\midrule[0.8pt]

$Q = \beta H (\rho_{\text{dm}} + \rho_{\text{de}})$
& $w_{\text{dm}}=0$; $w_{\text{de}}=-1$
& $72.68\pm0.90$                              & $0.359\pm0.020$                                     & $-0.120\pm0.044$                             & -- \\

$Q = \beta H (\rho_{\text{dm}} + \rho_{\text{de}})$
& $w_{\text{dm}}=0$; Free $w_{\text{de}}$
& $72.42\pm0.90$ &                         $0.181^{+0.061}_{-0.170}$                                       & $0.22^{+0.27}_{-0.18}$ &                                  $w_{\text{de}} = -0.758^{+0.170}_{-0.062}$ \\

$Q = \beta H (\rho_{\text{dm}} + \rho_{\text{de}})$
& Free $w_{\text{dm}}$; $w_{\text{de}}=-1$
& $72.44\pm0.91$                           & $0.333^{+0.021}_{-0.039}$                                     & $-0.39^{+0.16}_{-0.37}$ &                                 $w_{\text{dm}} = 0.23\pm0.24$ \\

\bottomrule[1.5pt]
\end{tabular}
\end{center}
\end{table*}

\begin{figure*}
\includegraphics[scale=0.7]{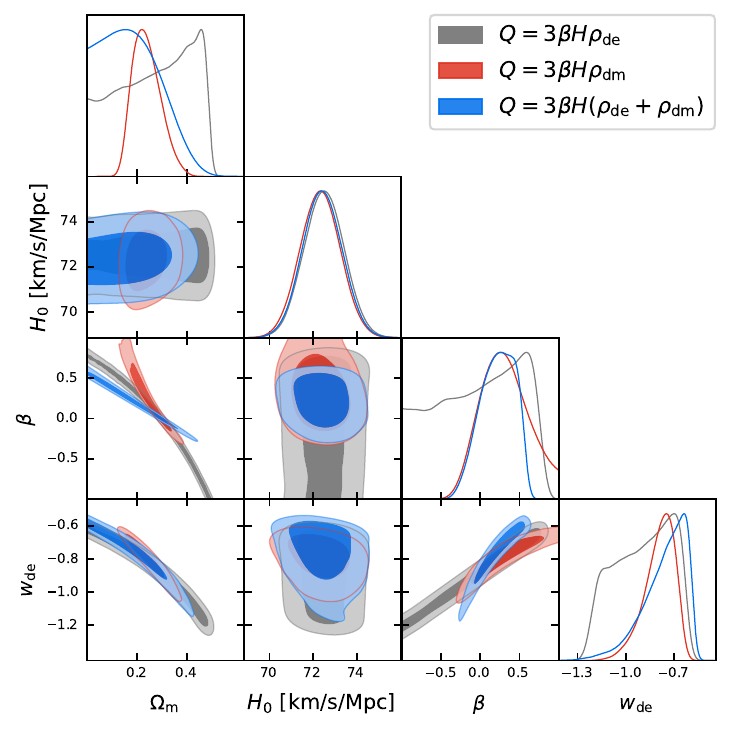}
\includegraphics[scale=0.7]{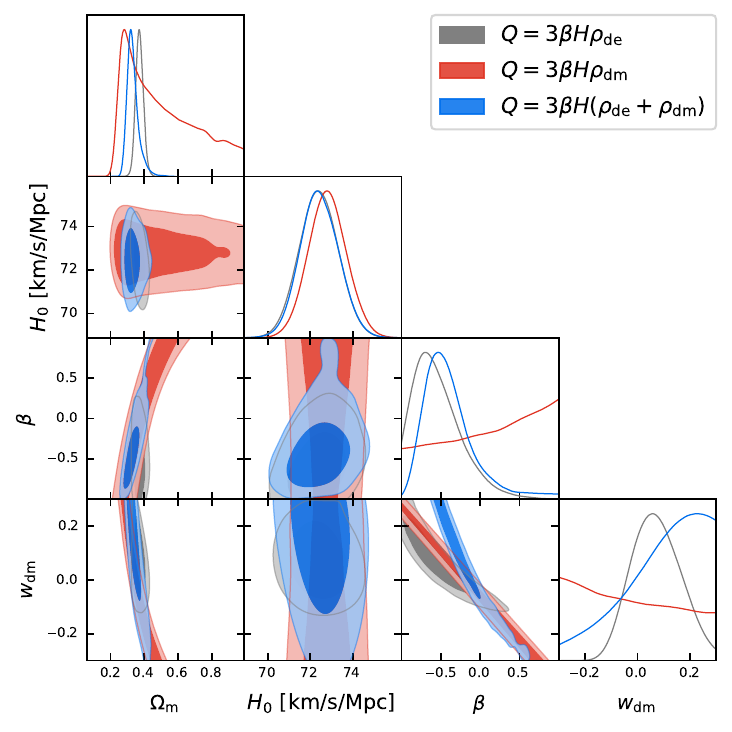}
\centering
\caption{Cosmological parameter constraints on IDE models from combined late-universe observations. Left: Constraints for IDE models with a free dark energy EoS while assuming cold dark matter ($w_{\text{dm}} = 0$). Right: Constraints for IDE models with a free dark matter EoS while fixing the cosmological constant ($w_{\text{de}} = -1$).}
\label{contours-3D}
\end{figure*}

\begin{figure*}
\includegraphics[scale=0.27]{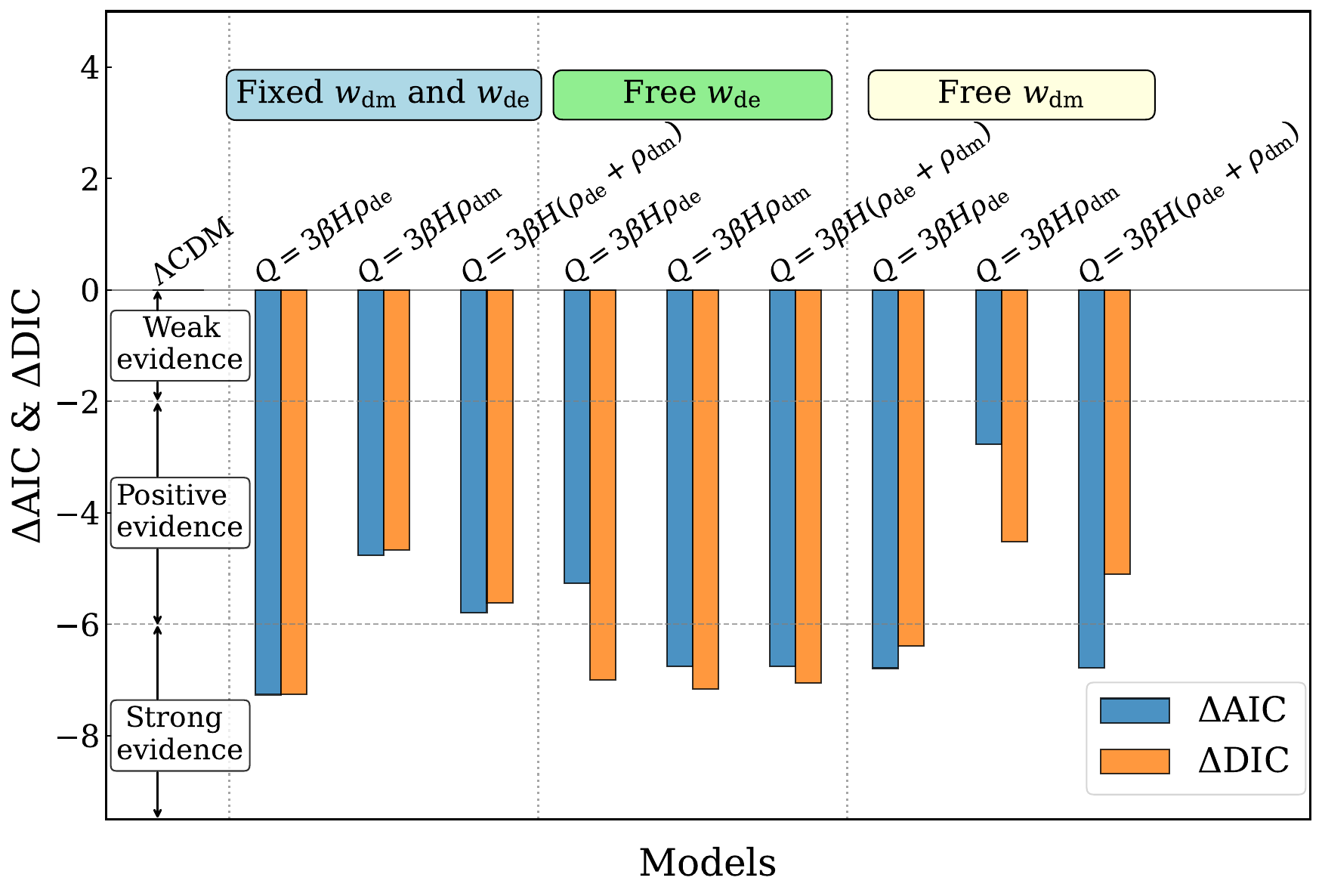}
\includegraphics[scale=0.27]{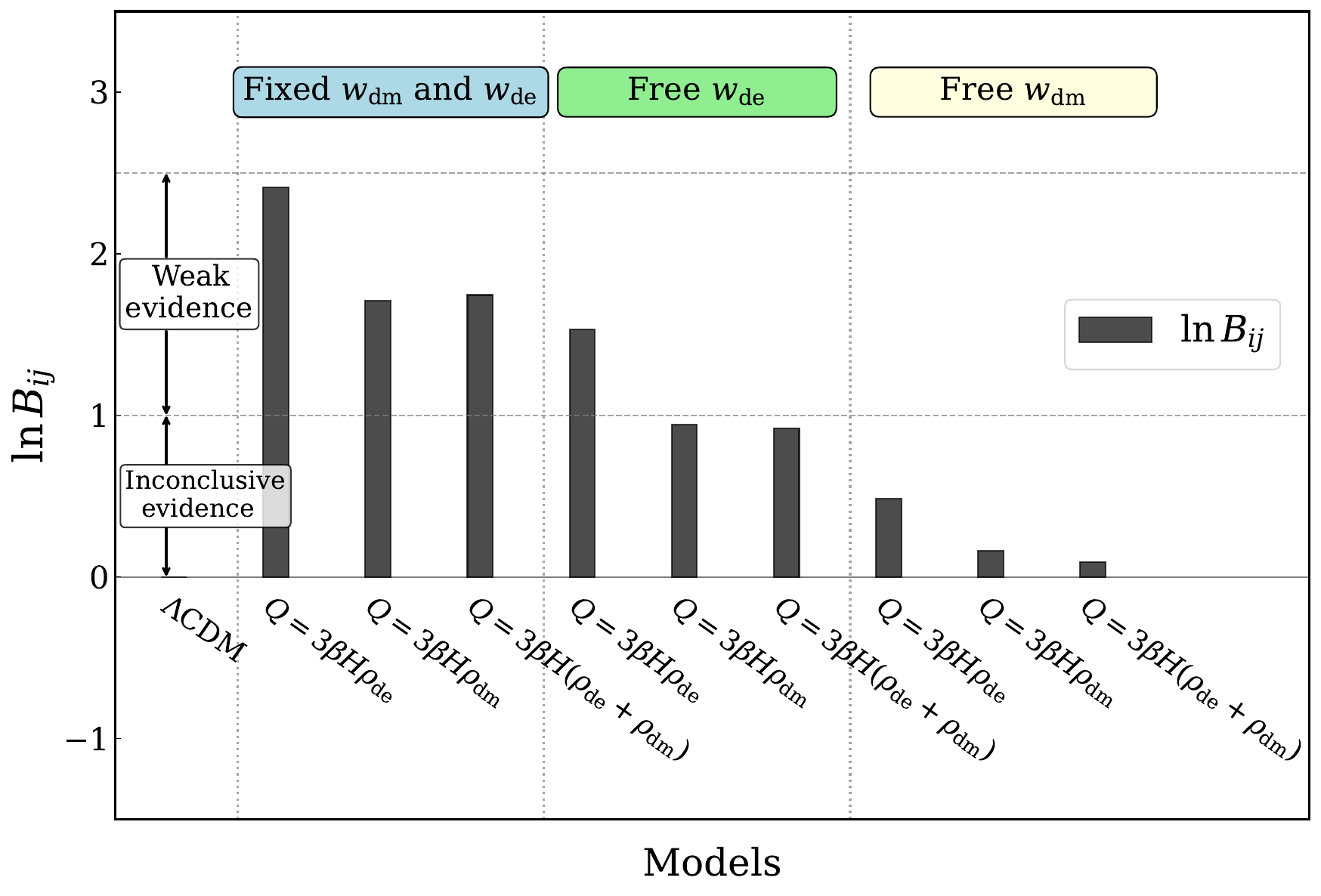}
\centering
\caption{Graphical representation of model comparison results. In this work, we choose the $\Lambda$CDM model as a comparison baseline. Left: The $\Delta$AIC and $\Delta$DIC values. A negative $\Delta$AIC/$\Delta$DIC value within the range $[-2,0)$ indicates weak evidence in favor of the model, a value within $[-6, -2)$ indicates positive evidence in favor, and a value within $[-10, -6)$ indicates strong evidence in favor. Right: The logarithmic Bayes factors. A positive \(\ln {\mathcal{B}}_{ij}\) value within the range $(0,1)$ indicates inconclusive evidence in favor of the model, and a value within $[1, 2.5)$ indicates weak evidence in favor of the model.}
\label{compare}
\end{figure*}

We adopt the MCMC method to infer the probability distributions of cosmological parameters, and conduct parameter sampling with {\tt Cobaya} \cite{Torrado:2020dgo}. The chain convergence is assessed via the Gelman-Rubin potential scale reduction factor ($R-1<0.01$ for all parameters). We fit the standard $\Lambda$CDM model, which serves as a comparison baseline, two models with free dark sector EoS, and IDE models to the data combination. All parameters are assigned flat priors, with ranges: $H_0\in [20, 100]\,\rm {km/s/Mpc}$, $\Omega_{\rm m}\in[0.01, 0.99]$, $\Omega_{\rm b}\in [0, 0.1]$, $r_{\rm d}\in[100, 200]\,\rm{Mpc}$, $w_{\rm dm}\in [-1, 1]$, $w_{\rm de}\in[-3, 1]$ and $\beta\in[-1, 1]$. In this analysis, $\Omega_{\rm b}$ is treated as a free parameter, and no external observational constraints are adopted as priors on either $\Omega_{\rm b}$ or the composite quantity $\Omega_{\rm b}h^2$. In general, adopting a Gaussian prior on $r_{\rm d}$ (e.g., from CMB observations) can help tighten constraints on cosmological parameters, but we forgo this treatment here to obtain pure low-redshift constraints. Notably, Verde et al. presented an almost model-independent determination of the standard ruler from low-redshift data ~\cite{Verde:2016ccp}. Such a measurement could potentially be used as a Gaussian prior on $r_{\rm d}$, offering a promising approach to tighten cosmological constraints. The $1\sigma$ (68.3\% confidence level) errors for the marginalized parameter constraints are summarized in Table~\ref{tab:results1}. To better elucidate the constraints on $\beta$ and its correlations with the dark energy EoS $w_\mathrm{de}$ and dark matter EoS $w_\mathrm{dm}$, Fig.~\ref{contours-3D} shows the constraint contours.

Our analysis begins by examining the non-interacting case ($Q=0$). When dark matter and dark energy EoS adhere to their standard values (\(w_{\text{dm}}=0\), \(w_{\text{de}}=-1\)), our constraints yield \(H_0 = 73.57\pm0.87 \ \rm{km/s/Mpc}\) and \(\Omega_{\rm m} = 0.3047\pm0.0075\). The $H_0$ constraint is consistent with the SH0ES result \(H_0 = 73.04\pm1.04\ \rm{km/s/Mpc}\). When allowing the dark energy EoS to deviate from \(-1\) while maintaining \(w_{\text{dm}}=0\), we observe \(w_{\text{de}} = -0.883^{+0.041}_{-0.037}\), which departs from the cosmological constant $\Lambda$ at a significance of $3.2\sigma$. Conversely, when allowing the dark matter EoS to deviate from \(0\) while fixing \(w_{\text{de}}=-1\), we obtain \(w_{\text{dm}} = -0.059^{+0.026}_{-0.022}\),  which departs from the cold dark matter case at a significance of $2.3\sigma$. These results therefore make it natural to explore the interplay between the dark sector interaction (governed by $Q$ and $\beta$) and the dark sector EoS. As an essential preliminary, we first establish how the introduction of interaction itself modifies the cosmological constraints. To this end, we examine the interacting scenarios under the canonical assumption of $w_{\text{dm}}=0$ and $w_{\text{de}}=-1$.

Under the assumption of cold dark matter and a cosmological constant, the constraints on IDE models reveal consistent patterns. For all three interaction forms considered, the coupling parameter $\beta$ is constrained to be negative, indicating a preference for energy transfer from dark energy to dark matter at $3.2\sigma$, $2.4\sigma$, and $2.7\sigma$ confidence levels, respectively. This interaction systematically affects the matter density parameter, with $\Omega_{\rm m}$ elevated to $0.378\pm0.024$, $0.347\pm0.018$, and $0.359\pm0.020$ for the three coupling forms respectively, compared to the $\Lambda$CDM value of $0.3047\pm0.0075$. This systematic increase in $\Omega_{\rm m}$ is consistent with the physical picture of energy transfer from dark energy to dark matter, as the transferred energy effectively increases the present-day dark matter density. We find that the constraints are dominated by BAO and SN, while adding CC, TD and GRB only slightly shifts $\beta$ toward zero and $\Omega_{\rm m}$ to lower values. We further find that the evidence for the interaction is weakened when using alternative SN datasets. Both results are detailed in Appendix~\ref{appendix:A}.

Allowing the dark energy EoS parameter $w_{\rm de}$ to deviate from its cosmological constant value, we observe systematic shifts in the constrained parameter spaces across all interaction forms. For $Q = \beta H\rho_{\rm de}$, freeing $w_{\rm de}$ causes the interaction parameter to move from significantly negative ($\beta = -0.35\pm0.11$) to near-zero ($\beta = -0.02^{+0.75}_{-0.38}$), while simultaneously driving $\Omega_{\rm m}$ to lower values ($0.378 \rightarrow 0.275$) and constraining $w_{\rm de} = -0.89^{+0.25}_{-0.13}$. For $Q = \beta H\rho_{\rm dm}$, the effect is even more pronounced: $\beta$ changes sign from negative ($-0.180^{+0.074}_{-0.067}$) to positive ($0.31^{+0.27}_{-0.35}$), $\Omega_{\rm m}$ decreases substantially ($0.347 \rightarrow 0.239$), and $w_{\rm de}$ converges to $-0.792^{+0.110}_{-0.063}$. The $Q = \beta H(\rho_{\rm dm}+\rho_{\rm de})$ form shows similar behavior with $\beta$ shifting from $-0.120\pm0.044$ to $0.22^{+0.27}_{-0.18}$ and $w_{\rm de}$ converges to $-0.758^{+0.170}_{-0.062}$. In summary, allowing for dynamical dark energy ($w_{\rm de} > -1$) resolves the apparent need for interaction. The data prefer a universe where the late-time acceleration is driven by dark energy with properties slightly different from a cosmological constant (quintessence-like behavior), rather than by energy transfer between the dark sectors. This preference is evidenced by the fact that the deviations of $\beta$ from zero are constrained at $0.03\sigma$, $0.9\sigma$, and $1.2\sigma$ confidence levels for the three interaction coupling forms, respectively, whereas the significance of $w_{\rm de}$ departing from $-1$ reaches substantially higher values of $0.8\sigma$, $3.3\sigma$, and $3.9\sigma$, respectively.

The dark matter EoS, when treated as a free parameter, systematically alters the inferred strength and direction of the dark sector interactions. This influence is clearly demonstrated across different coupling forms. For $Q = \beta H\rho_{\rm de}$, we find $w_{\rm dm} = 0.071^{+0.088}_{-0.100}$ correlating with $\beta = -0.53^{+0.16}_{-0.39}$, reinforcing the preference for energy transfer from dark energy to dark matter. In stark contrast, for $Q = \beta H\rho_{\rm dm}$, the constraint $w_{\rm dm} = -0.20^{+0.20}_{-0.49}$ necessitates a sign reversal in the interaction, yielding $\beta = 0.42^{+1.50}_{-0.54}$ (i.e., dark matter decays into dark energy). For the symmetric coupling $Q = \beta H(\rho_{\rm dm}+\rho_{\rm de})$, the constraint $w_{\rm dm} = 0.23\pm0.24$ correlates with $\beta = -0.39^{+0.16}_{-0.37}$, maintaining the same anti-correlation pattern observed in the asymmetric case. Across different coupling forms, we observe a consistent pattern: a positive $w_{\rm dm}$ correlates with energy transfer from dark energy to dark matter ($\beta<0$), while a negative $w_{\rm dm}$ pairs with energy transfer in the opposite direction ($\beta>0$). This compensatory relationship demonstrates that the cosmological effects of dark matter pressure can be equivalently described by a suitable interaction between the dark sectors. In contrast to the case with a free $w_{\mathrm{de}}$, when $w_{\mathrm{dm}}$ is freed, the data continue to favor an interaction. The parameter $\beta$ deviates from zero at $3.3\sigma$, $0.8\sigma$, and $2.4\sigma$ confidence levels for the three interaction forms, respectively. In comparison, the deviation of $w_{\mathrm{dm}}$ from zero is substantially less significant, at $0.7\sigma$, $1.0\sigma$, and $0.96\sigma$ confidence levels, respectively.

As clearly illustrated in Fig.~\ref{contours-3D}, $\beta$ is positively correlated with $w_{\rm de}$ but negatively correlated with $w_{\rm dm}$. Notably,  the degeneracy between $w_{\rm de}$ and $\beta$ is the strongest for the interaction form $Q = \beta H \rho_{\text{de}}$, whereas the degeneracy between $w_{\rm dm}$ and $\beta$ is most pronounced for $Q = \beta H \rho_{\text{dm}}$. This dependence indicates that the interaction structure strongly impacts the resulting parameter constraints, highlighting the critical need to investigate which interaction form is most physically motivated and consistent with observational data.

Our parameter constraints on the Hubble constant demonstrate consistency with local measurements, with $H_0$ values ranging between $72.33$ and $72.81$ km/s/Mpc across all IDE scenarios, aligning with the SH0ES collaboration's measurement of $73.04\pm1.04$ km/s/Mpc. The tight $H_0$ uncertainties indicate that the dataset can strongly constrain the current expansion rate regardless of the interaction mechanism. Freeing the dark energy EoS consistently yields lower $\Omega_{\rm m}$ values across all interaction models, indicating a robust degeneracy where a more negative $w_{\text{de}}$ compensates for a reduced matter density. In contrast, freeing the dark matter EoS shows no universal trend: $\Omega_{\rm m}$ may increase or decrease depending on the specific interaction form, highlighting the model-dependent nature of $w_{\text{dm}}$ and its complex role in balancing the cosmological energy budget.

Having established the parameter constraints for IDE models, we perform a model comparison using the Akaike information criterion (AIC) and deviance information criterion (DIC) to assess their statistical support from the observational data. These criteria balance model fit and complexity, defined as:
\begin{align}
\text{AIC} = \chi^2_{\rm min} + 2k; \ \ \ \text{DIC} = \chi^2_{\rm min} + 2k_{\rm eff},
\end{align}
where $k$ is the number of free parameters, and $k_{\rm eff} = \langle \chi^2\rangle - \chi^2_{\rm min}$ denotes the number of effectively constrained parameters, with the angular brackets indicating the average over the posterior distribution. Models yielding smaller AIC/DIC values are considered to be more favored. We are not concerned with the absolute AIC/DIC values, but the relative ones between different models, $\Delta{\rm AIC}/\Delta{\rm DIC}$. Here, we adopt $\Lambda$CDM as the comparison baseline. Therefore, negative $\Delta{\rm AIC}/\Delta{\rm DIC}$ values denote superior fit to data relative to $\Lambda$CDM, whereas positive values indicate inferior fit. Generally, a $\Delta{\rm AIC}/\Delta \rm{DIC}$ value within $[-2, 0)$ indicates weak evidence in favor of the model, a value within $[-6, -2)$ denotes positive evidence in favor, and a value within $[-10, -6)$ denotes strong evidence in favor. Conversely, the interpretation for positive values is symmetric, indicating equivalent levels of evidence against the model. Our statistical analysis reveals that all IDE models receive observational support relative to the $\Lambda$CDM model, with evidence strengths ranging from positive to strong, as summarized in the left panel of Fig.~\ref{compare}. This consistent pattern reinforces the viability of IDE scenarios in explaining the cosmological observations.

To enhance the robustness of our conclusions, we further adopt the Bayes factor for model comparison. The Bayes factor is derived from Bayesian evidence, which is defined as the marginal likelihood of the data \(D\) for a model \(M\):
\begin{equation}
Z = \int_{ \Omega} P(D|{\bm \theta}, M) P({\bm \theta}|M) P(M){\rm d}{\bm \theta},
\end{equation}
where \( P(D|\boldsymbol{\theta}, M) \) is the likelihood, \( P(\boldsymbol{\theta}|M) \) is the parameter prior, and \( P(M) \) is the model prior. The logarithmic Bayes factor (\(\ln {\mathcal{B}}_{ij}\)) for comparing models \(i\) and \(j\) is given by \(\ln {\mathcal{B}}_{ij} = \ln Z_i - \ln Z_j\), with \(Z_i\) and \(Z_j\) representing the Bayesian evidence of models \(i\) and \(j\), respectively. For the interpretation of \(\ln {\mathcal{B}}_{ij}\): support for model \(i\) over model \(j\) is inconclusive if \(0<\ln \mathcal{B}_{ij} <1\), weak if \(1\leq \ln \mathcal{B}_{ij} <2.5\), moderate if \(2.5\leq \ln \mathcal{B}_{ij} <5\), and strong if \(\ln\mathcal{B}_{ij}\ge5\); conversely, a negative \(\ln \mathcal{B}_{ij}\) indicates evidence against model \(i\). Our results show that the Bayes factor only provides inconclusive to weak support for IDE models compared to the $\Lambda$CDM model, as summarized in the right panel of Fig.~\ref{compare}, which leaves ample room for \(\Lambda\)CDM to remain viable. Among these models, those with $Q = \beta H \rho_{\text{de}}$ are mildly favored, those with the dark energy and dark matter EoS fixed to their $\Lambda$CDM values are more favored, and those with a free dark energy EoS are preferred over those with a free dark matter EoS.

\section{Conclusion}\label{sec4}
This work presents an investigation of IDE scenarios using late-universe observations, examining how freeing dark energy and dark matter EoS affect the strength and direction of dark sector interactions.

Our analysis reveals the following key findings. When the dark sector EoS are fixed to $w_{\text{de}}=-1$ and $w_{\text{dm}}=0$, all interaction forms consistently yield negative coupling parameters $\beta$, indicating a preference for energy transfer from dark energy to dark matter. However, this apparent evidence is critically dependent on the EoS assumptions. We identify significant degeneracies between $\beta$ and dark sector EoS. When $w_{\text{de}}$ is allowed to deviate from $-1$, the strong evidence for interaction substantially weakens across all coupling forms. This systematic pattern demonstrates that the observational signatures previously attributed to dark sector interaction can be well explained by a dark energy with $w_{\text{\rm de}} > -1$. In contrast, when $w_{\mathrm{dm}}$ is freed, the data continue to favor an interaction between dark sectors. We find that the positive $w_{\mathrm{dm}}$ correlates with energy transfer from dark energy to dark matter ($\beta < 0$), while negative $w_{\mathrm{dm}}$ associates with energy flow in the opposite direction ($\beta > 0$). Generally, $\beta$ is positively correlated with $w_{\rm de}$ but negatively correlated with $w_{\rm dm}$, and the interaction form strongly impacts the resulting constraints. This motivates further studies to identify the physically favored interaction form. In addition, the model comparison using both AIC and DIC shows that all IDE models considered receive substantial statistical support over the $\Lambda$CDM cosmology, with evidence levels ranging from positive to strong. However, the Bayesian evidence analysis provides only inconclusive to weak support for IDE models, which leaves ample room for $\Lambda$CDM to remain viable.

Despite these advances, our analysis has several limitations. Notably, we have not explored scenarios where $w_{\rm de}$ and $w_{\rm dm}$ are simultaneously freed, nor have we considered the time-evolving EoS. These choices are primarily due to the limited constraining power of current late-universe datasets, which would yield uninformatively large uncertainties if applied to such extended parameter spaces. Moving forward, we plan to pursue several promising avenues. First, we will investigate the inclusion of CMB data, which could help break degeneracies through its complementary sensitivity to early-universe physics and thereby impose tighter constraints on the interplay between interactions and the properties of dark matter and dark energy. While promising, this approach requires careful consideration of the potential tensions between CMB and late-time observations regarding the interaction direction, as presented in Ref.~\cite{Li:2024qso}. Second, we will examine how synergies of emerging late-universe probes \cite{Wu:2022dgy,Wu:2022jkf} can improve constraints on these more complex interacting scenarios.

\begin{acknowledgments}
We thank Jian-Hua Yang, Guo-Hong Du and Tian-Nuo Li for fruitful discussions. We thank the anonymous referee for constructive comments on this work. This work was supported by the Research Starting Funds for Imported Talents, Ningxia University (Grant No. 030700002562). Shang-Jie Jin is grateful for the support from the China Scholarship Council, and JST ASPIRE Program of Japan (Grant No. JPMJAP2320).
\end{acknowledgments}

\section{Data Availability}
The data that support the findings of this paper are openly available \cite{DESI:2025zgx,DES:2024jxu,Moresco:2022phi,Wu:2025wyk,Amati:2018tso}.

\bibliography{IDE-refs}

\appendix
\newpage
\section{Cosmological Constraints from Different Probe Combinations}\label{appendix:A}
In this appendix, we investigate the impacts of individual cosmological probes on the IDE constraints, as well as the effects of adopting different SN datasets. We take the IDE model with $w_{\text{de}} = -1$, $w_{\text{dm}} = 0$ and $Q=\beta H \rho_{\rm de}$ as our illustrative example,  which is the most favored model in this work based on current late-universe observations (see Fig.~\ref{compare}).

\begin{figure*}[htbp!]
\centering
\includegraphics[scale=0.35]{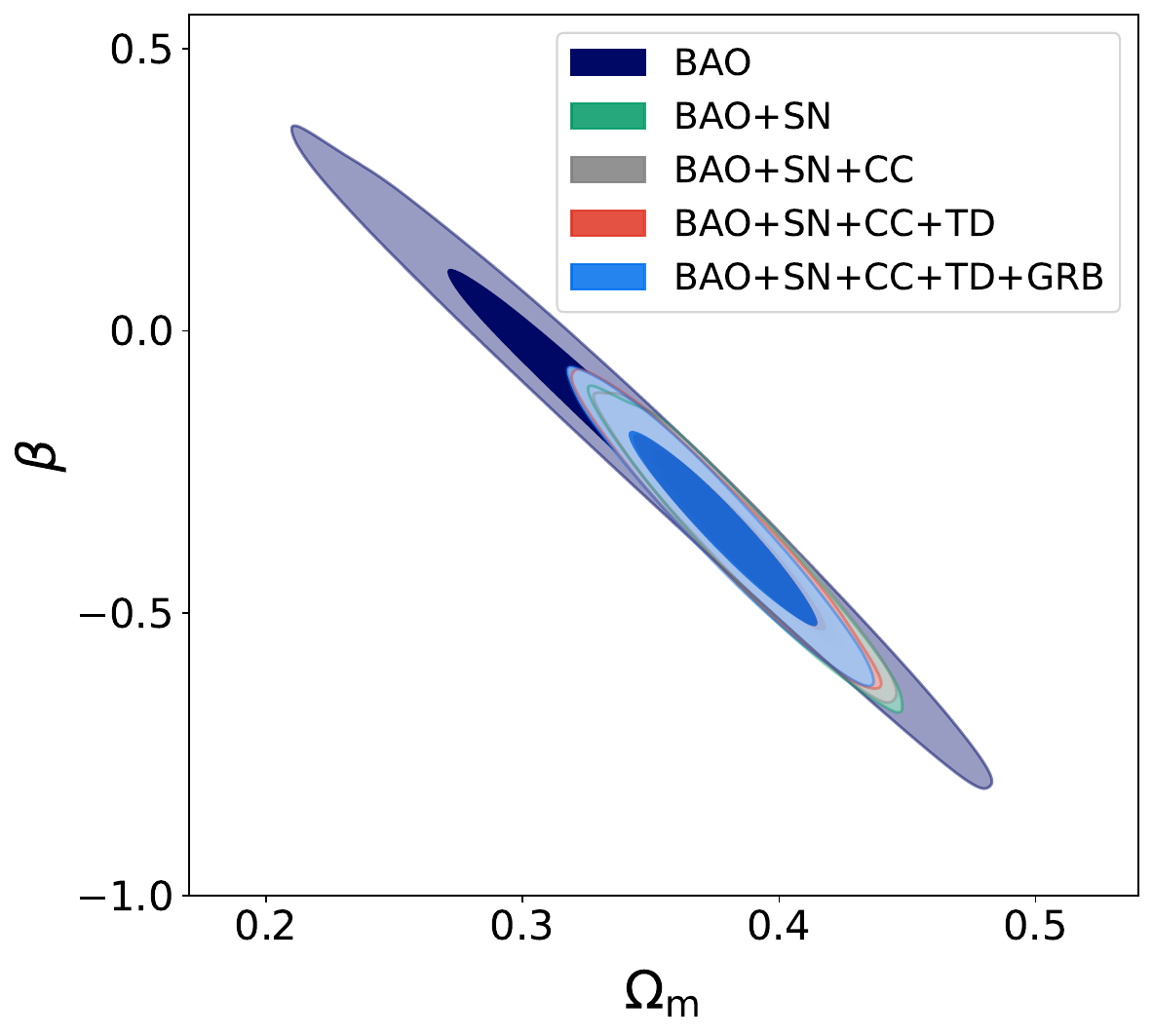}
\includegraphics[scale=0.35]{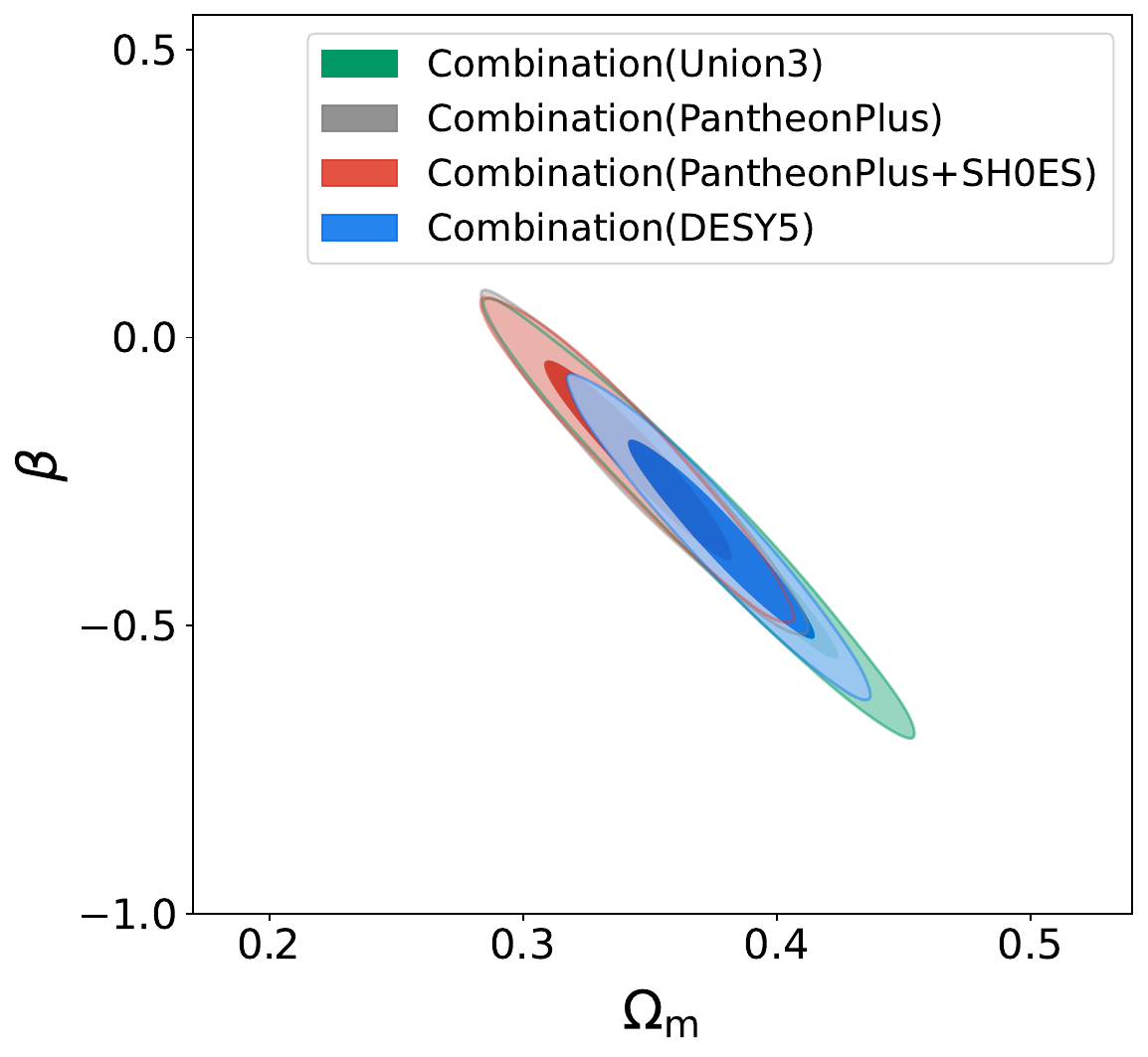}
\caption{Cosmological parameter constraints on the IDE model with $w_{\text{de}} = -1$, $w_{\text{dm}} = 0$ and $Q=\beta\rho_{\rm de}$ from combined late-universe observations. Left: Constraints from sequential combinations: BAO, BAO+SN, BAO+SN+CC, BAO+SN+CC+TD, and BAO+SN+CC+TD+GRB. Right: Constraints from the full combination, using different SN datasets (DESY5, Union3, PantheonPlus, and PantheonPlus+SH0ES) while keeping all other probes identical.}
\label{probes}
\end{figure*}

We first adopt a sequential combination approach to assess the contribution of each probe, using the order: BAO, BAO+SN, BAO+SN+CC, BAO+SN+CC+TD, and BAO+SN+CC+TD+GRB. As shown in the left panel of Fig.~\ref{probes}, the constraints are dominated by BAO and SN, while adding CC, TD, and GRB yields only a modest improvement in precision, due to their limited data quantity and quality. Individually, the inclusion of CC, TD, or GRB slightly shifts the coupling parameter $\beta$ toward zero, corresponding to a weaker indication of the dark sector interaction. These probes also marginally shift $\Omega_{\rm m}$ toward smaller values. The analysis is based on the latest SN sample, DESY5 \cite{DES:2024jxu}. It is well known that different SN datasets can lead to distinct cosmological parameter constraints. For instance, the official DESI DR2 analysis found that, for the $w_0w_a$CDM model, the significance of the rejection of $\Lambda$CDM depends on the SN sample used. When combining DESI and CMB with PantheonPlus \cite{Brout:2022vxf}, Union3 \cite{Rubin:2023jdq}, and DESY5, the rejection significance is $2.8\sigma$, $3.8\sigma$, and $4.2\sigma$, respectively, compared to $3.1\sigma$ for DESI+CMB \cite{DESI:2025zgx}. Motivated by this, we next explore the effects of adopting different SN datasets on the IDE constraints.

In our analysis, we marginalize over the absolute magnitude $M_B$ for three SN datasets, whereas some studies instead marginalize over the degenerate combination of $M_B$ and $H_0$ (see Ref.~\cite{DES:2024jxu}). Importantly, marginalizing over $M_B$ alone is mathematically equivalent to marginalizing over the $M_B$--$H_0$ combination for constraining other cosmological parameters, ensuring the robustness of our conclusions. The DESY5 analysis acknowledges the inherent $M_B$--$H_0$ degeneracy and refrains from constraining $H_0$ to avoid unphysical results. In contrast, we treat $H_0$ as a free parameter with a flat prior $H_0 \in [20, 100]\,\rm{km/s/Mpc}$ and only marginalize over $M_B$. This choice allows joint constraints on $H_0$ with other probes such as CC and TD within a unified framework. We find that different SN datasets notably affect the constraint results: PantheonPlus and Union3 tend to weaken the preference for dark sector interactions, with Union3 yielding weaker constraints than PantheonPlus and DESY5. Specifically, the joint analysis yields results of $\beta=-0.32\pm0.15$ for Union3 and $\beta=-0.22\pm0.12$ for PantheonPlus, respectively. Additionally, calibrating $M_B$ via SH0ES does not substantially alter constraints on dark sector interactions ($\beta=-0.21\pm0.12$), despite improving the determination of $H_0$, a quantity not central to this work and thus not discussed further here.

\end{document}